\documentclass[11pt]{article}

\usepackage{natbib,amsmath,amssymb,amsthm,amsfonts}
\usepackage{graphicx,url}
\usepackage{booktabs}
\usepackage{latexsym}
\usepackage[stable]{footmisc}
\usepackage{denselists}
\usepackage{authblk}
\usepackage{hyperref}
\usepackage{cleveref}

\usepackage{xcolor}

\newcommand{\Id}{\operatorname{Id}}

\newcommand{\bo}{\mathbf{1}}
\newcommand{\bz}{\mathbf{0}}

\DeclareMathOperator{\diag}{diag}

\newcommand{\sighat}{\hat{\Sigma}}

\theoremstyle{plain}

\title{Comparison of REML methods for the study of phenome-wide genetic variation}

\author[1,2]{Damian Pavlyshyn}
\author[1]{Iain M.\ Johnstone}
\author[3]{Jacqueline L.\ Sztepanacz}
\affil[1]{Department of Statistics, Stanford University}
\affil[2]{Disease Elimination Program, Burnet Institute}
\affil[3]{Department of Ecology and Evolutionary Biology, University of Toronto}

\date{October 2022}

\begin{document}
\maketitle

\begin{abstract}
   It is now well documented that genetic covariance between functionally related traits leads to an uneven distribution of genetic variation across multivariate trait combinations, and possibly a large part of phenotype-space that is inaccessible to evolution. How the size of this nearly-null genetic space translates to the broader phenome level is unknown. High dimensional phenotype data to address these questions are now within reach, however, incorporating these data into genetic analyses remains a challenge. Multi-trait genetic analyses, of more than a handful of traits, are slow and often fail to converge when fit with REML. This makes it challenging to estimate the genetic covariance (\textbf{G}) underlying thousands of traits, let alone study its properties. We present a previously proposed REML algorithm that is feasible for high dimensional genetic studies in the specific setting of a balanced nested half-sib design, common of quantitative genetics. We show that it substantially outperforms other common approaches when the number of traits is large, and we use it to investigate the bias in estimated eigenvalues of \textbf{G} and the size of the nearly-null genetic subspace. We show that the high-dimensional biases observed are qualitatively similar to those substantiated by asymptotic approximation in a simpler setting of a sample covariance matrix based on i.i.d. vector observation, and that interpreting the estimated size of the nearly-null genetic subspace requires considerable caution in high-dimensional studies of genetic variation. Our results provide the foundation for future research characterizing the asymptotic approximation of estimated genetic eigenvalues, and a statistical null distribution for phenome-wide studies of genetic variation.    
\end{abstract}

\section{Introduction}
\label{sec:backgr-motiv}

To answer the most compelling questions in evolutionary biology we must uncover the causal connection between genotypes, phenotypes, and the environment (selection). Unlike the finite genomes that underlie them, organisms are comprised of essentially infinite phenotypes that may genetically vary and covary. While not all of those phenotypes will be meaningful for fitness, there are at least many thousands to consider, if we are to begin to understand the genotype-phenotype map. Efforts to increase the scope and throughput of phenotyping have been cited as an urgent priority in evolutionary biology since at least 2010 \citep{houle_numbering_2010, houle_phenomics_2010}. High-throughput phenomic technologies such as such as RNA sequencing \citep{schrag_beyond_2018}, drone based plant imaging \citep{furbank_phenomicstechnologies_2011}, wearable sensors \citep{haleem_biosensors_2021, sharma_wearable_2021, neethirajan_recent_2017}, metabolomics for physiological measurements \citep{jin_genetic_2020, freimer_human_2003}, and computer vision for morphometrics \citep{lurig_computer_2021} are now within reach. However, incorporating these high-dimensional phenotype data into genetic analyses remains a challenge. 

Multivariate studies of genetic variation focusing on relatively small sets of traits  $p \ll 20$ (small \textit{p}) have transformed our understanding of how genetic variation is distributed across phenotypes, and how this affects evolutionary outcomes. While almost all individual traits studied have been shown to have genetic variation \citep{lynch_genetics_1998}, and responses to artificial selection are often rapid and of large magnitude \citep{hill_what_2010}, multivariate studies of the genetic variance-covariance matrix (\textbf{G}) show that this genetic variation is distributed unevenly across multivariate phenotypes \citep{kirkpatrick_patterns_2009, sztepanacz_crosssex_2019}. The concentration of genetic variance onto fewer multivariate trait combinations than the number of phenotypes measured is biologically caused by the pleiotropic effects of alleles on multiple traits which leads to their genetic covariance \citep{lande_genetic_1980}.  Multivariate trait combinations with high genetic variation form a genetic subspace where traits are predicted to have high evolvability. Evolution is predicted to occur more quickly along these multivariate trait combinations than in any individual trait \citep{agrawal_how_2009}, contributing to divergence among populations (eg. \citep{mcglothlin_conservation_2022, schluter_adaptive_1996}), species (eg. \citep{innocenti_interspecific_2013, begin_micro-_2004}), and sexes (eg. \cite{gosden_evolutionary_2014}). The set of orthogonal multivariate trait combinations with low genetic variation, form a complementary subspace which is termed the nearly-null genetic subspace \citep{gomulkiewicz_demographic_2009, gaydos_visualizing_2013}. This subspace putatively represents important evolutionary constraints in natural populations, where phenotypic evolution is expected to be constrained \citep{gomulkiewicz_demographic_2009}, occur slowly \citep{kirkpatrick_patterns_2009} or stochastically \citep{hine_evolutionary_2014}.

Estimating the size of the nearly-null genetic subspace is an avenue to quantify both genetic constraints that may lead to evolutionary limits, and the extent of pleiotropy underlying organisms, which determines their genetic dimensionality.  Past studies have typically found that genetic variance is restricted to less than half of the phenotype space with most multivariate trait combinations having no detectable genetic variation \citep{blows_distribution_2015} . One notable exception to this pattern is \textit{Drosophila} wing shape, which as been shown to have genetic variation in all multivariate wing shape traits (ie. a full rank \textbf{G}) \citep{mezey_dimensionality_2005, sztepanacz_dominance_2015, sztepanacz_crosssex_2019, houle_estimating_2015}. These multivariate studies have typically dealt with small sets of functionally related traits such as wing \citep{mezey_dimensionality_2005}, skeletal \citep{garcia_quantitative_2014} or phyto \citep{walsh_evolution_2018} morphology, or traits that have a strong physiological \citep{caruso_genetic_2005}, or biochemical relationships \citep{sztepanacz_reduced_2012}. Therefore, genetic covariance among these traits may be relatively high compared to the range of traits that are possible to study with phenomics. Whether inferences for the size of the nearly null subspaces found in these small \textit{p} studies can be extrapolated to phenomes is an open biological question.  

Confounded with the biological phenomena that nearly-null genetic
subspaces represent, are the statistical phenomena that arise from
their estimation. Even in low \textit{p} studies, estimating
\textbf{G} and its eigenvalues is a challenge. The standard approach
is to fit a multivariate mixed model in REML to estimate \textbf{G} as
an unstructured covariance matrix. These models commonly fail to
converge with only a small number of traits, and run-times are long
because of the quadratic growth $\sim p^2/2$ in the number of
parameters to estimate with the number of traits. As shown in the
one-way MANOVA context, small \textit{p} models that do converge,
often produce estimated matrices that are not positive definite
\citep{hill_probabilities_1978} and that have overdispersed
eigenvalues \citep{hayes_modification_1981}, properties which are
exacerbated in the two-way hierarchical model, such as the nested
half-sib design common in quantitative genetic studies. In particular, \citet{hayes_modification_1981} show that the magnitude
of overdisperion of the genetic eigenvalues is determined by
$p/n$
with smaller sample sizes or more traits leading to larger
dispersion.  The overdispersion of estimated eigenvalues is a pattern
remarkably similar to that we interpret biologically as genetic
subspaces of high evolvability and those that are nearly null. In the
phenomic context, where
$p/n \sim 1$ or possibly even greater than 1,
it could even obscure any biological signal. Disentangling how much of the dispersion of eigenvalues of \textbf{G} is due to biological covariance versus statistical estimation is therefore important, if we want to predict evolutionary constraints.

In simpler settings, recent work has provided a quantitative
description of the bias and error related to the dispersion of
estimated eigenvalues
when $p$ is of order comparable to $n$,
summarized for example in  \citep{jopa18}. The bulk eigenvalue distribution of sample (i.i.d)
covariance matrices, such as those describing among-individual
covariance structure such as phenotypic covariance matrices
\textbf{P}, or genomic relatedness matrices (GRMs), is known to follow
the Marchenko Pastur distribution
\citep{mapa67,yzb15}.
For a large class of mixed models, including full-sib and half-sib
designs, and for the particular case of MANOVA estimators,
overdisperson of sample eigenvalues in the bulk eigenvalue
distribution is described by a generalization of the Marchenko-Pastur
distribution \citep{fan_eigenvalue_2019}.

The individual eigenvalues of sample covariance matrices also follow
particular distributions. 
The leading eigenvalue conforms to the Tracy-Widom distribution,
both for sample (i.i.d.) covariance matrices \citep{john00c}
  and for MANOVA estimates from mixed models\citep{fajo22}. The trailing (smallest) eigenvalue converges to a reflected
Tracy-Widom distribution \citep{fajo22}.
Again for MANOVA estimates in mixed models, the leading
estimated eigenvalues and eigenvectors can be influenced by an
alignment between
directions of sufficiently elevated variance in the
different covariance components \citep{fan_spiked_2018}. For example,
an alignment between the major axis of genetic variance 
$\mathbf{g}_{\it max}$
and environmental variation could lead to an estimated $\mathbf{g}_{\it max}$ that is biased toward the major axis of environmental variation and with an eigenvalue that is much larger than it should be. These quantitative descriptions provide appropriate null distributions for MANOVA eigenvalues, and enable bias-correction for estimated eigenvalues in these contexts. 

Simple among line (one-way) genetic simulation studies estimating \textbf{G} for \textit{p}=5 using REML, combined with an empirical centering and scaling approach \citep{saccenti_tracy-widom_2011}, suggest that REML eigenvalues may have similar properties. The bulk distribution of eigenvalues appears to follow the Marchenko Pastur distribution \citep{blows_distribution_2015}, and the leading eigenvalues of estimated \textbf{G} are consistent with the Tracy Widom distribution \citep{sztepanacz_accounting_2017}. Notably, however, the leading eigenvalues of \textbf{G} estimated using factor analysis or MCMCglmm do not follow the Tracy Widom distribution \citep{sztepanacz_accounting_2017}, showing that the method used to estimate \textbf{G} influences the sampling distribution of its eigenvalues. These studies laid the groundwork for determining the properties of REML eigenvalues in the one-way design, however, they are limited to small \textit{p} and $n \gg p$. A major hurdle in extrapolating these studies to the phenomic-scale, is the difficulty in employing REML for high \textit{p}.

Alternative approaches to REML have been developed in recent years and employed to estimate \textbf{G} and its eigenvalues for high \textit{p} phenomic studies. 
The Bayesian sparse factor approach of 
\citet{runcie_dissecting_2013}  enables the estimation of genetic covariance for thousands of traits by placing a prior on the latent factors underlying \textbf{G} and \textbf{E} that few traits contribute to them (ie. they are sparse). Studies using this method have qualitatively found the same uneven distribution of standing genetic variation \citep{garcia_quantitative_2014}, mutational variation \citep{hine_uneven_2018}, and a large nearly-null genetic subspace, as seen in small \textit{p} studies. While biologically motivated, the assumption that sparse factors underlie both \textbf{G} and \textbf{E} may upwardly bias estimates of the nearly null subspace.  MegaLMM \citep{runcie_megalmm_2021} uses strong Bayesian priors on the sparsity and number of important latent factors underlying \textbf{G} and \textbf{E} to fit linear mixed models for $>20,000$ traits and relatively small \textit{n}. The primary goal of this approach, however, is for genomic prediction, not estimation. Consequently, the properties that make the method feasible for including phenome-level data, and which are beneficial for genomic prediction, may have undesirable properties with respect to estimation. 
\citet{blows_phenome-wide_2015} circumvented convergence problems in REML by constructing large \textbf{G} from a set of overlapping principal sub-matrices, estimated for a small numbers of traits using REML. Since the constructed matrix is not guaranteed to be positive definite, a subsequent bending \citep{hayes_modification_1981, meyer_bending_2019} of \textbf{G} was required. This approach enabled a coarse description of the distribution of genetic variation across \textbf{G}, finding that the vast majority of genetic variation was found in few dimensions. However, a quantitative description of the size of the nearly null subspace was not possible. While each of these approaches have value in certain circumstances, they all make some undesirable assumptions that lead to challenges in interpreting the size of the nearly null space and performing statistical hypothesis testing.

In this paper we demonstrate that REML is feasible for high \textit{p} phenomics in the specific setting of a balanced nested half-sib design, without having to make any additional assumptions about the structure of \textbf{G} or perform any bending, and we begin to study the properties of estimated eigenvalues from these high \textit{p} models.  In particular, we perform simulation study investigating the spectral characteristics of the REML estimate of $\mathbf{G}$ for a family of balanced half-sib breeding designs in the moderate\nobreakdash-$p$ ($p\approx 50$) setting.

As algorithms that are usually used to fit mixed models
often fail to converge even for $p$ as small as $5$, we use an algorithm designed specifically for fitting balanced nested designs introduced in \cite{cady91} that will allow us to find REML estimates for larger values of $p$.

In \cref{sec:initial-examples}, we compare this Calvin-Dijkstra algorithm with various other procedures used to compute or approximate the REML estimate, showing that it does at least as well as its common alternatives.
We then demonstrate that the algorithm is a practical method for REML estimation for the balanced designs even for $p\approx 50$, in which regard it substantially outperforms the alternatives presented.

Sections \ref{sec:repeated-sampling} and \ref{sec:nearly-null-subsp}
  investigate the illustrate the biases inherent in using functions of eigenvalues of the REML estimate $\hat{\mathbf{G}}$ to estimate the corresponding functions of the eigenvalues of $\mathbf{G}$.
We show that, as $p$ gets large, the large eigenvalues of $\hat{\mathbf{G}}$ have a substantial upward bias when used to estimate the large eigenvalues of $\mathbf{G}$.
We also investigate the bias in estimators of the nearly-null dimension of $\mathbf{G}$ based on counting small eigenvalues of $\hat{\mathbf{G}}$.
These biases are more difficult to describe, so we demonstrate the
behavior of such estimators on a family of qualitatively different
between-sire and between-dam covariance matrices.
Section \ref{sec:conclusions} has some summary conclusions and discussion.

\section{Description of methods compared}
\label{sec:descr-meth-comp}

Our simulations use a balanced half-sib (or nested two-way) design
\begin{equation}
  \label{eq:model-a}
Y_{i j k}=\mu+\alpha_i+\beta_{i j}+\varepsilon_{i j k}
\end{equation}
Here $\alpha_i, \beta_{i j}$ and $\varepsilon_{i j k}$ are the random
effects due to sire, dam and individual respectively, and $\mu$ is a
fixed intercept.

Each $Y_{ijk}$ is a vector of $p$ traits; there are $I, J$ and $K$
possible values of the indices $i,j$ and $k$.
Each random effect vector is assumed to follow independent normal
distributions
\begin{equation}
  \label{eq:model-b}
    \alpha_i \sim \mathcal{N}(0, \Sigma_A), \qquad
  \beta_{ij} \sim \mathcal{N}(0, \Sigma_B), \quad
  \epsilon_{ijk}\sim \mathcal{N}(0, \Sigma_E).
\end{equation}

We are particularly interested in estimation of the genetic covariance
matrix $\textbf{G} = \Sigma_A$ and its eigenvalues, but will also look at the
concomitant estimates of $\Sigma_B$ and $\Sigma_E$. The estimates we
will compare are 

\begin{itemize}
\item MANOVA, which are easy to compute but do not necessarily produce
  a positive-definite estimate for $\Sigma_A$, 
\item REML-\textsf{lme}. fitted using a generic mixed-effects model
  solver (we use \texttt{R} function \texttt{lme()}), which should converge to
  the true REML estimates but are prohibitively slow to compute for
  high (or even moderate) numbers of traits, 
\item pseudo-REML, as described by \citet{amem85}. These
  are easy to compute, and in the full-sub (one-way) design are
  exactly the REML estimates. In the half-sib case, they are only
  approximations. 
\item stepwise REML, as described by \citet{cady91}
  This is an iterative algorithm that converges to the true
  REML estimates. Although we know of no results decribing
  the rate of convergence, in practice it is very fast, allowing for
  easy REML estimation even for $100 \times 100$ matrices. 
\item pairwise REML, in which the estimates of the individual entries
  $\Sigma_{lm}$ are computed using a bivariate analysis with traits
  $l$ and $m$. This estimate is sometimes proposed as a computational
  fall-back when the dimensionality of $\Sigma$ is too high for REML
  to converge; it need not be non-negative definite, as will be seen.
\end{itemize}

\section{Initial examples
}
\label{sec:initial-examples}

\subsection{A small \texorpdfstring{$p$}{p} example with zero eigenvalues
}
\label{sec:small-q-example}

As a warm-up, we first illustrate the methods in a setting with a
small number of traits, $p = 4$.\footnote{indeed, this was the largest number
  of traits for which \texttt{lme()} runs relatively quickly.}
Specifically, we take
\begin{equation}
  \label{eq:par-a}
  p=4, \qquad
  I = 100, \qquad
  J = 3, \qquad
  K = 5,
\end{equation}
and diagonal structures for the variance component matrices:
\begin{alignat}{2}
  \label{eq:par-b}
  \Sigma_A
  & = \sigma_A^2 \diag(1,1,0,0),  \qquad \qquad & \sigma_A^2 & = 25 \notag \\
  \Sigma_B
  & = \sigma_B^2 \diag(1,0,0,1), & \sigma_B^2 & = 9, \\
  \Sigma_E
  & = \sigma_E^2 \Id_4, & \sigma_E^2 & = 1. \notag
\end{alignat}
This information is not used when fitting -- all procedures are run
with no assumptions on the covariances.
The vector $\mu = (1,2,3,4)$, but its value is immaterial to all
methods used here.

All simulations were done in \texttt{R}; the software and technical descriptions of code used to produce the results of this paper are available in an \texttt{R} package hosted at \url{https://github.com/damian-t-p/REMLSimulationPaper}.

Table \ref{tab:four-traits} shows the results of one sample
realization, about which the following remarks may be made.
\begin{itemize}
\item For MANOVA, as expected, the zero eigenvalues of $\Sigma_A$ lead
  to some negative estimated eigenvalues. The REML criterion is higher
  than that of the actual REML estimate found by stepwise REML. This
  is because MANOVA in this balanced setting in fact maximizes the
  likelihood over a larger set, i.e. without enforcing the
  non-negativity constraints.
\item the Calvin-Dykstra (CD) algorithm for stepwise-REML exhibited
  linear convergence in the REML criterion (not shown here). The CD
  convergence criterion uses
  \[
  d^2(\hat{\Sigma}^l, \hat{\Sigma}^{l-1})
  := \sum_{k \in \{A, B, E\}} n_k \lVert\hat{\Sigma}_k^l - \hat{\Sigma}_k^{l-1}\rVert_2^2,
\]
where $\hat{\Sigma}^l = (\hat{\Sigma}^l_A, \hat{\Sigma}^l_B, \hat{\Sigma}^l_E)$ is the vector of covariance estimates after the $l$th iteration.
We stopped when $d(\hat{\Sigma}^l,
\hat{\Sigma}^{l-1}) < 10^{-6}$. 
\item The pseudo REML estimates
  are close, but not identical, to the
  stepwise REML ones, and the REML criterion score is lower.
\item REML-\textsf{lme} is slow and often fails to converge: we record
  the results at termination, even before convergence, and for this
  reason the REML criterion score is less than that of stepwise REML,
  which does converge to the maximum.
\end{itemize}




\begin{table}[h]
  \centering
\begin{tabular}{llrrrrc}
\toprule
Component  & Method & $\lambda_1$ & $\lambda_2$ & $\lambda_3$ & $\lambda_4$ & REML-Crit. \\
\midrule
$\Sigma_A$ & MANOVA        & $26.24$ & $24.62$ & $-0.01$ & $-0.13$         & $-7039$ \\
           & REML-LME      & $26.27$ & $24.46$ & $0.34$  & $4.3\text{e-}3$ & $-7083$ \\
           & Stepwise REML & $26.24$ & $24.61$ & $0$     & $0$             & $-7040$ \\
           & Pseudo REML   & $26.24$ & $24.62$ & $0$     & $0$             & $-7054$ \\
\midrule
$\Sigma_B$ & MANOVA        & $9.35$ & $8.72$ & $0.015$         & $-2.5\text{e-}3$ \\
           & REML-LME      & $9.17$ & $8.47$ & $0.025$         & $0.015$          \\
           & Stepwise REML & $9.31$ & $8.65$ & $7.8\text{e-}3$ & $0$              \\
           & Pseudo REML   & $9.35$ & $8.74$ & $0.015$         & $0$              \\
\midrule
$\Sigma_E$ & MANOVA        & $1.10$ & $1.00$ & $0.93$ & $0.93$ \\
           & REML-LME      & $1.09$ & $1.01$ & $0.96$ & $0.92$ \\
           & Stepwise REML & $1.10$ & $1.00$ & $0.93$ & $0.93$ \\
           & Pseudo REML   & $1.10$ & $1.00$ & $0.93$ & $0.93$ \\
\bottomrule
\end{tabular}
  \caption{Sample eigenvalues $\lambda_1 \geq \lambda_2 \geq \lambda_3
  \geq \lambda_4$ for one sample realization drawn from
  \eqref{eq:model-a}--~\eqref{eq:model-b} with parameters as in \eqref{eq:par-a}--~\eqref{eq:par-b}.
 }
  \label{tab:four-traits}
\end{table}



\subsection{An example with \texorpdfstring{$p = 50$}{p = 50}
}
\label{sec:an-example-with}

An important advantage of stepwise REML is that it is can easily be
run for relatively large numbers of traits. 
In this section, we use $p = 50$, but $p$ can be in the thousands
before the stepwise REML
estimate takes more than a second to compute. 

The parameters are as in \eqref{eq:par-a}--~\eqref{eq:par-b}, but with
some modifications:
\begin{equation}  \label{eq:q50}
  p = 50, \qquad
  \Sigma_A = \sigma_A^2 \diag ( |Z_i|), \qquad
  \Sigma_B = \sigma_A^2 \diag ( |Z_i'|),
\end{equation}
where $Z_i, Z_i'$ are all independent standard normal deviates.

\begin{figure}[htbp]
  \centering
  \includegraphics[width=.8\textwidth]{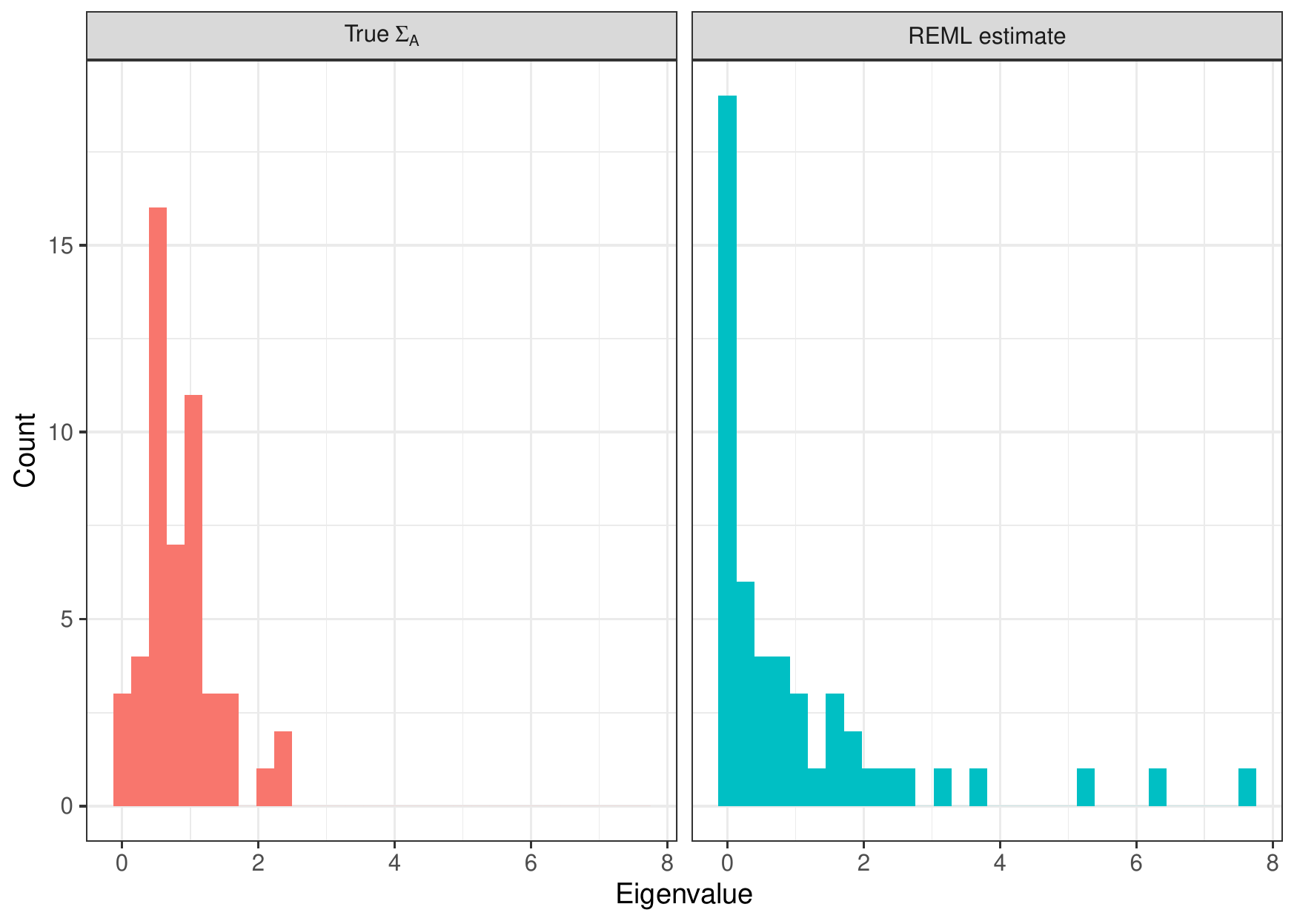}
  \caption{Histograms of eigenvalues of $\Sigma_A$ and its REML estimate}
  \label{fig:q50}
\end{figure}

Figure \ref{fig:q50} compares the eigenvalues of $\hat{\Sigma}_A$ (in
blue) next to the true eigenvalues of $\Sigma_A$ (in red).
Now that $p = 50$ is of the same order as $I = 100$, the sample
eigenvalues are much more spread out than those of $\Sigma_A$ -- 
we see that the REML estimate zeroes out many eigenvalues while
overestimating a few others.


\subsection{Pairwise REML with \texorpdfstring{$p = 50$}{p = 50}}
\label{sec:pairwise-reml}

We investigate the strategy of computing pairwise REML estimates
for each pair of traits. 
We create a new estimate $\hat{\Sigma}_{A,P}$ such that the diagonal
entries $[\hat{\Sigma}_{A,P}]_{ii}$ are computed by doing a
single-trait analysis on trait $i$ and the off-diagonal entries
$[\hat{\Sigma}_{A,P}]_{ij}$ are computed by doing a 2-trait analysis
on the traits $(i,j)$. 

The setting is the same as the previous subsection, especially
\eqref{eq:q50}. 
In fact, we used the stepwise-REML estimates for the 2-trait analyses,
since there are 1250 entries of $\hat{\Sigma}_{A,P}$ to fill in, and even
in the 2-dimensional case, fitting a generic mixed model is too slow
for this. 
However, we can still compare the resulting $\hat{\Sigma}_{A,P}$ with
the REML esimate $\hat{\Sigma}_{A,R}$ to see how the eigenvalues
produced by the pairwise procedure differ from those done by computing
the whole stepwise REML estimate at once.

\begin{figure}[htbp]
  \centering
  \includegraphics[width=.8\textwidth]{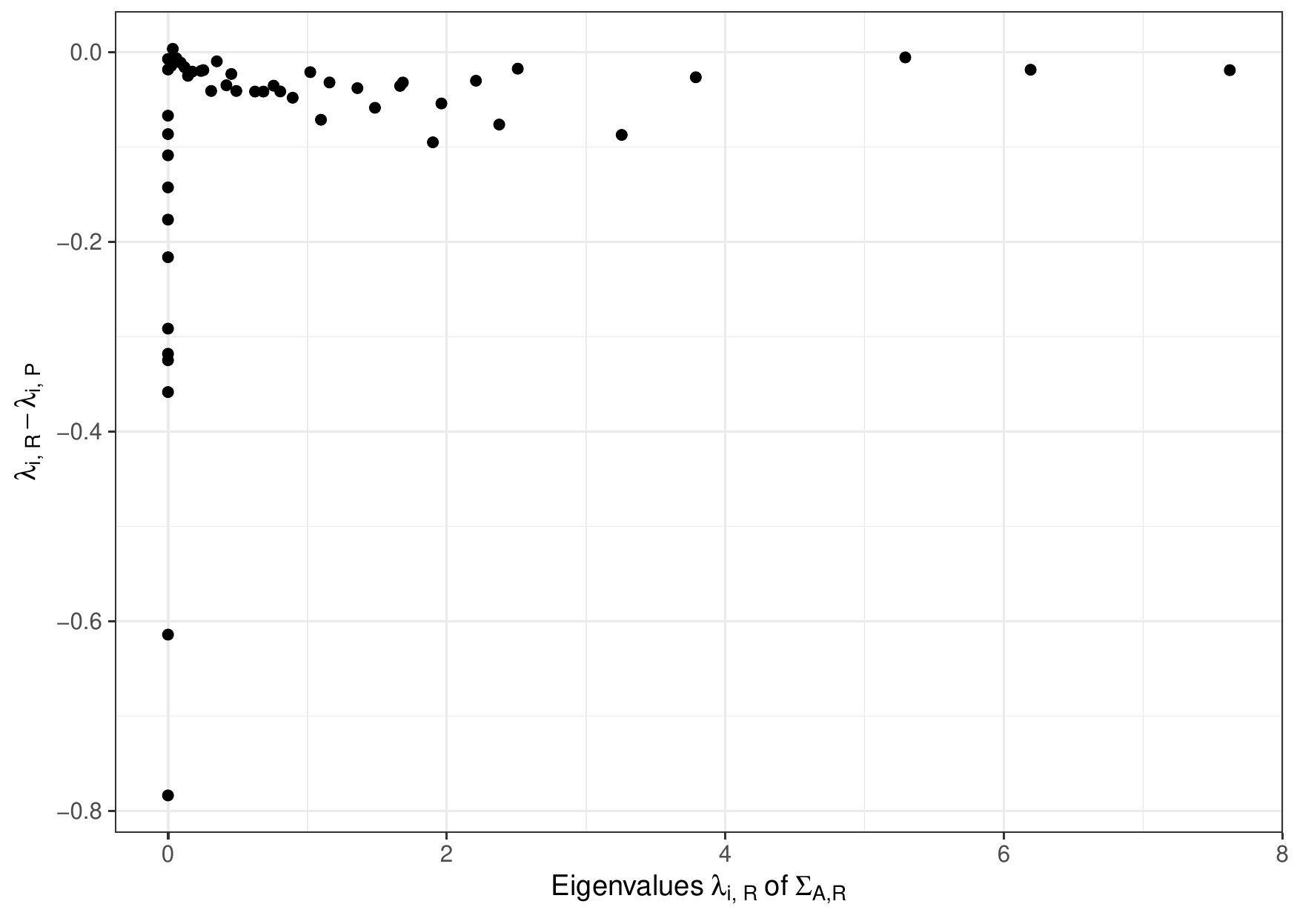}
  \caption{Differences $\lambda_{i,R} - \lambda_{i,P}$ of eigenvalues of $\Sigma_{A,R}$ and $\Sigma_{A,P}$ plotted against $\lambda_{i,R}$.}
  \label{fig:p-REML}
\end{figure}

Let $\lambda_{1,R} > \ldots > \lambda_{p,R}$ be the eigenvalues of
$\hat{\Sigma}_{A,R}$ and
$\lambda_{1,P} > \dotsc > \lambda_{p,P}$ be the eigenvalues of $\hat{\Sigma}_{A,P}$. 
Figure \ref{fig:p-REML} plots the differences 
$\lambda_{i,R} - \lambda_{i,P} $
against $\lambda_i$.
We may observe that
\begin{itemize}
\item the pairwise REML estimate $\hat{\Sigma}_{A,P}$ is not positive
  definite: some $15$ of its estimated eigenvalues $\lambda_{i,P}$ lie between $0$ and
  $-0.75$,
\item the pairwise REML eigenvalues are all smaller than the actual
  REML ones: \quad $\lambda_{i,P} \leq \lambda_{i,R}$ for all $i$,
\item when the pairwise REML eigenvalues are positive, the differences
  are not large: an empirical summary (in this case) might be that typically,
  \begin{equation*}
    0 \leq \lambda_{i,R} - \lambda_{i,P} \leq 0.05 \lambda_{i,R}
  \end{equation*}
\end{itemize}


\section{Repeated sampling}
\label{sec:repeated-sampling}


The Calvin-Dykstra algorithm is fast enough that it is possible to
study the properties of REML estimates of eigenvalues through simulation.




\subsection{Bias in top eigenvalues \texorpdfstring{$\lambda_{1,R}, \ldots, \lambda_{5,R}$}{lambda(1,R), ..., lambda(5,R)}
}
\label{sec:bias-top-eigenvalues}

This example
confirms the upward bias in the largest eigenvalues
of the REML estimates and confirms that the bias increases with the
number of traits $p$.

Again we use the half-sib design
\eqref{eq:model-a}--~\eqref{eq:model-b} with $I=100, J=3, K=5$ and
\begin{equation}
  \label{eq:lg-ev}  
  \Sigma_A = I, \qquad
  \Sigma_B = 4 I, \qquad
  \Sigma_E = I.
\end{equation}
The number of traits $p$ grows from 10 to 100 in increments of 10.
Over $R = 50$ replications, the averages of the top 5 eigenvalues of
$\sighat_{A,R},\sighat_{B,R}$ and $\sighat_{E,R}$ are plotted against
$p$.
Also shown (thick red lines) are the corresponding population
eigenvalues ($1,4$ and $1$ respectively).

\begin{figure}[htbp]
  \centering
  \includegraphics[width=.8\textwidth]{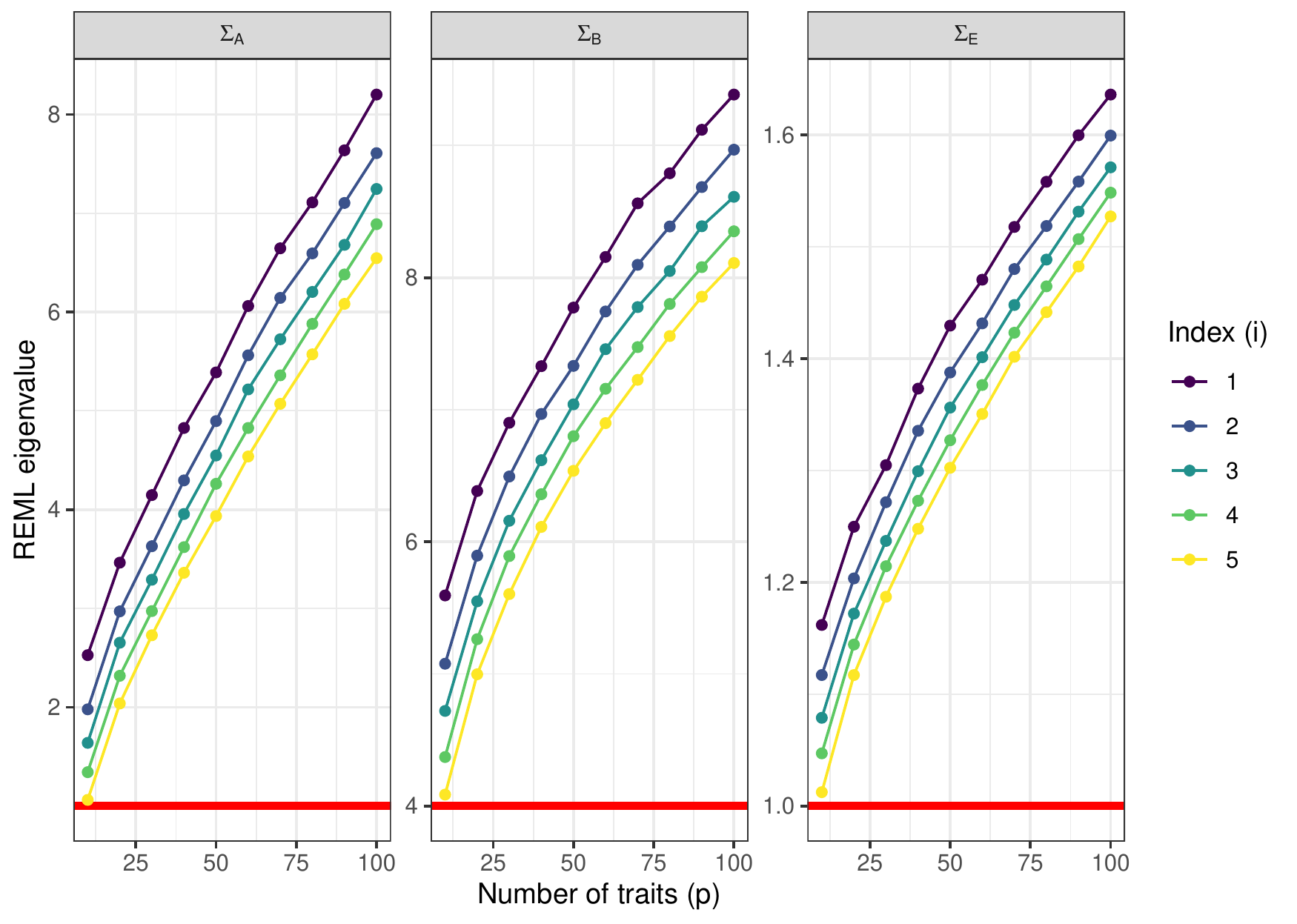}
  \caption{The largest 5 eigenvalues of the REML estimates of $\Sigma_A, \Sigma_B$ and $\Sigma_E$ plotted against $p$.
  Each point is the mean of $50$ replicates.
  The largest population eigenvalue of each matrix is depicted with a horizontal red line.}
  \label{fig:largest}
\end{figure}

Some remarks on the results in Figure \ref{fig:largest}:
\begin{itemize}
\item the means are biased upwards, as expected, but the biases are
  much more substantial for $\Sigma_A$ (especially) and $\Sigma_B$
  than for $\Sigma_E$.
  Also notable is the rapid increase in the biases with $p$ in each case,
\item The standard deviations of the 50 replicates were computed for
  each eigenvalue (results not shown here).
  These were markedly larger for $\sighat_{A,R}$ and
  $\sighat_{B,R}$ (between 0.2 and $0.35$)
  than for $\sighat_{E,R}$ (between 0.02 and 0.03).
  Although obscured by sample fluctuation, it appears --- as might be
  expected --- that the SDs are ordered, decreasing as we move down
  from the top eigenvalue to the 5th. There is no clear dependence on
  $p$ in the SDs.
\item In this setting the REML estimates coincide with the MANOVA estimators. This
  is because all of the population eigenvalues of $\Sigma_A, \Sigma_B$
  and $\Sigma_E$ in \eqref{eq:lg-ev} are well separated from $0$, 
  and the MANOVA estimates were in all cases positive definite.
\end{itemize}

\subsection{Differences between the largest REML and MANOVA eigenvalues}
\label{sec:comparing}

In contrast to the previous example, when nearly null genetic
subspaces are present we expect the REML estimates to differ
significantly from MANOVA as the smallest MANOVA eigenvalues will be
negative.
Here we examine the difference between the \textit{largest} eigenvalues
$\lambda_i(\sighat_{k,R})$ and $\lambda_i(\sighat_{k,M})$ for REML and
MANOVA for $k \in \{A, B, E\}$.

\begin{figure}[htbp]
  \centering
  \includegraphics[width=.8\textwidth]{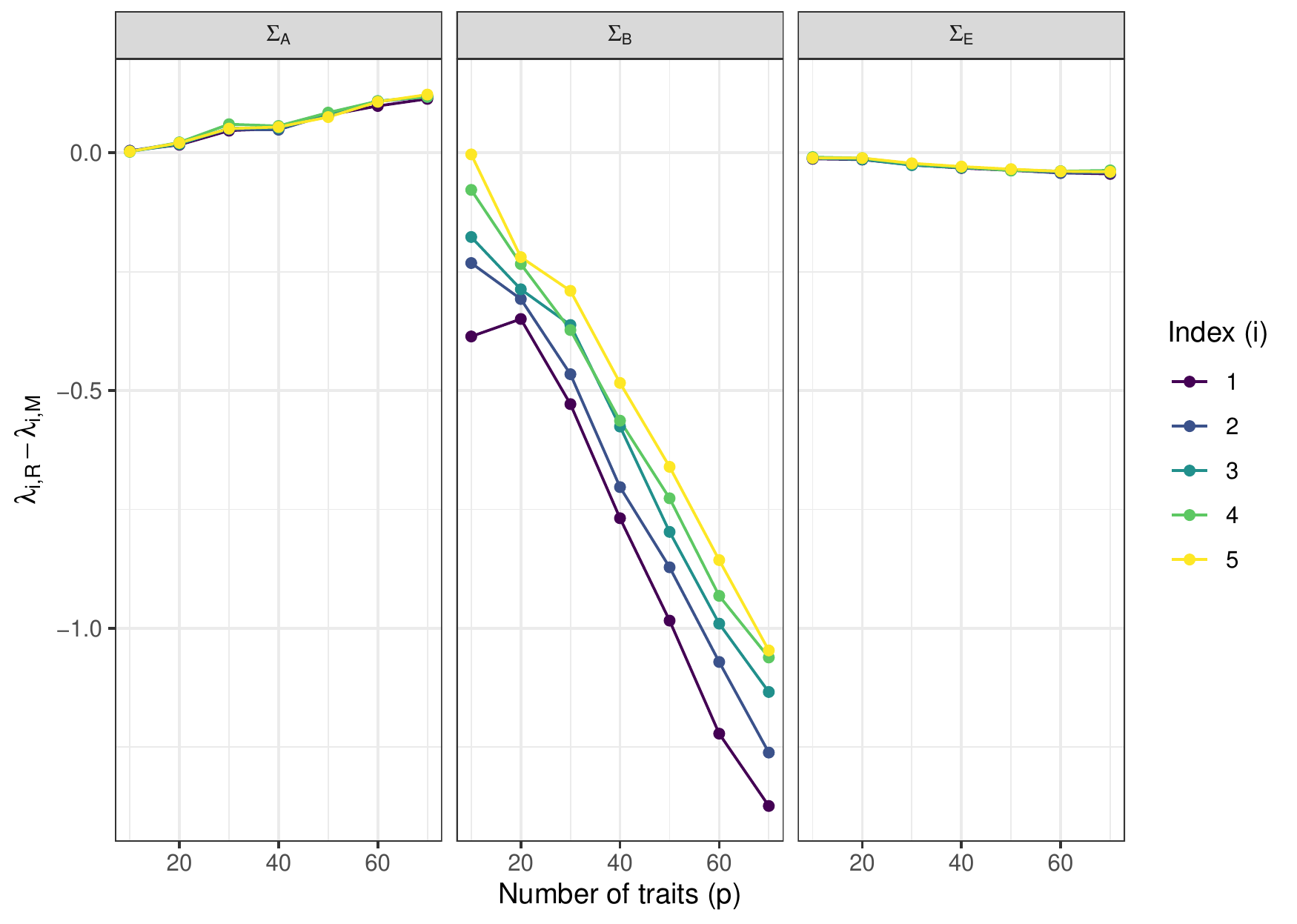}
  \caption{Differences between the REML and MANOVA estimates of the largest 5 eigenvalues of $\Sigma_A, \Sigma_B$ and $\Sigma_E$ plotted against $p$.
  Each point is the mean of $50$ replicates.}
  \label{fig:e-diff}
\end{figure}

We set half of the diagonal entries of $\Sigma_A$ and $\Sigma_B$ to
zero at random, leaving $\Sigma_E$ and all other parameters unchanged
from the previous subsection.
Figure \ref{fig:e-diff} shows the mean differences (averaged over 50
replications) of $\lambda_i(\sighat_{k,R}) - \lambda_i(\sighat_{k,M})$
as a function of $p = 10, 20, 30, \dotsc,  100$. We observe that
\begin{itemize}
\item the mean differences again increase in
  absolute value, nearly linearly,  as $p$ increases,
\item for $\Sigma_A$ the top REML eigenvalues are
  larger than MANOVA, while for $\Sigma_B$ and $\Sigma_E$ they are
  smaller,
\item the magnitude of the (negative) difference is much larger for
  $\Sigma_B$,
\item the SDs of the differences over the 50 replications (not shown) are
  approximately 10\% of the mean differences, and in particular are
  generally increasing with $p$.
\end{itemize}
These variations in sign and magnitude of the differences between the
top REML and MANOVA eigenvalues call for further study. At the least
they indicate a dependence on the level of nesting of the respective
variance components.

\section{Nearly Null Subspaces}
\label{sec:nearly-null-subsp}

\subsection{Counting zero eigenvalues in REML estimator \texorpdfstring{$\hat{\Sigma}_{A,R}$}{Sigma\^(A,R)}
}
\label{sec:count-zero-eigenv}

In this section we simulate several half-sib designs in which
$\Sigma_A$  has null spaces of various dimensions $d$.
We might expect the number of zero (or small) eigenvalues of the
corresponding REML estimates $\sighat_{A,R}$ to track
the null space dimension $d$.
We will see that this is broadly true, but that again significant
biases are present.

\medskip
\textit{Specification of $\Sigma_A, \Sigma_B, \Sigma_E$ -- generalities}. \quad
In general, the principal axes of variation (i.e. eigenvectors) of the
covariance matrices $\Sigma_A, \Sigma_B, \Sigma_E$ need not be
aligned, and so to use only diagonal matrices in our investigations
would entail a loss of generality.

We will therefore allow for examples of non-alignment of the axes of 
$\Sigma_A$ and $ \Sigma_B$, but for simplicity we keep the error
covariance scalar: $\Sigma_E = I$.
Because the model \eqref{eq:model-a}---~\eqref{eq:model-b} is
Gaussian, the REML and MANOVA estimates are unchanged by rotation (an
orthogonal transformation) of the data.
This means that we can take $\Sigma_A$ to be diagonal (and $\Sigma_E =
I$ is unchanged), but that $\Sigma_B$ need not be.

Specifically suppose that $\Sigma_A$ has eigendecomposition $\Sigma_A
= P \Lambda P^T$ for suitable orthogonal and diagonal matrices $P$ and
$\Lambda$ respectively.
Then the REML estimates are the same for the model with parameters
$(\Sigma_A, \Sigma_B, \Sigma_E = I)$ as for the model with
parameters $(\Lambda, P^TBP, I)$. 

In summary, the following are important to note:
\begin{Itemize}
\item We can only assume a diagonal structure for one of $\Sigma_A$ or
  $\Sigma_B$,   but not both,
\item If we are interested in estimating something about $\Sigma_A$ other
  than the eigenvalues, we cannot assume a diagonal structure for
  $\Sigma_A$, 
\item If we do not assume that $\Sigma_E$ is a scalar matrix, then it cannot
  be assumed to be diagonal either. 
\end{Itemize}

\medskip
\textit{Settings for $\Sigma_A$:} 
Thus, we will consider the following specifications for $\Sigma_A$:
\begin{align}
  \Sigma_A
   & =
   \begin{cases}
     c_A \diag( \bo_{p-d}, \bz_d)   & \text{Identity} \\
     c_A \diag( (X_i)_1^{p-d}, \bz_d) \qquad \quad X_i \stackrel{\rm
       iid}{\sim} \chi_5^2  & \text{Chi-squared} \\
     c_A \diag( (|X_i+5|)_1^{p-d}, \bz_d)  \quad X_i \stackrel{\rm
       iid}{\sim} {\textrm{Cauchy}}(0)  & \text{Heavy-tail}
   \end{cases}
\end{align}
with the choices
\begin{equation*}
  p  = 50,  \qquad d = 0, 5, 10, \dotsc, 50, \qquad c_A = 0.5, 1, 2.
\end{equation*}

A typical draw from each of these cases (with $d = 10$ zeros) might be
summarized qualitatively as follows:
\begin{itemize}
\item Identity:
  40 non-zero eigenvalues all equal to 1,
\item Chi-squared: 40 nonzero eigenvalues spread unevenly over the range
  $(0,15)$,
\item Heavy-tailed:  39 nonzero eigenvalues spread unevenly over $(0,30)$ with
  an outlier (e.g. at 750).
\end{itemize}

\medskip
\textit{Settings for $\Sigma_B$:}  We consider one scalar and three
non-diagonal cases:
\begin{equation}
    \Sigma_B  =
    \begin{cases}
      I & \text{`Identity'} \\
      W_q(q, I)/q & \text{`Wishart'} \\
      P \diag( (X_i)_1^q) P' \quad X_i \stackrel{\rm
       iid}{\sim} \chi_5^2,  \quad P \ \textrm{random orthogonal}
      & \text{`High-rank'} \\
      0.8 \bo \bo' + 0.2 I  & \text{`High-corr'}
    \end{cases} 
\end{equation}

\bigskip
\bigskip
\vspace{3em}
In a little more detail:
\begin{itemize}
\item Identity:
  sets all 50 eigenvalues  equal to 1,
\item Wishart: Let $X$ be a $p \times p$ matrix of iid $N(0, 1)$
  random variables. Then we form $B = XX^{\mathrm{T}}/p$, which is the
  sample covariance of $X$. Since the population covariance of the
  columns of $X$ is $I$, $B$ is not too far from $I$, while still
  having uniformly distributed eigenvectors. The eigenvalues of $B$
  are unevenly spread between 0 and 4\footnote{and are approximately a sample
  from a Marcenko-Pastur quarter circle law with parameter $\gamma =
  1$}.
\item High-rank: Let $P$ be a matrix of orthonormal vectors drawn
  uniformly. Let $D$ be diagonal matrix with iid $\chi^2_5$ entries on
  the diagonal. We then take $B = PDP^{\mathrm{T}}$. This matrix has
  eigenvectors independent from those of $A$, but the eigenvalues are
  more spread out than those in the Wishart case.
\item High-corr: $B$ is a square matrix with $1$s on the diagonal and
  $0.8$s in every other entry. This can be represented as $B = 0.8
  \mathbf{1}\mathbf{1}^{\mathrm{T}} + 0.2 I$, where $\mathbf{1}$ is a
  vector of ones. This means that the eigenvalues of $B$ are $0.8 p = 40$
  with multiplicity $1$ and $0.2$ with multiplicity  $p-1 = 49$. 
\end{itemize}

\medskip
\textit{Remaining parameter settings:}   In all cases, the error covariance is
taken to be the identity $\Sigma_E = I$.
We use the following set parameter values:
\begin{itemize}
\item Number of traits, $p = 50$, 
\item Number of sires, dams per sire, and individuals per dam: $I = 100, J
= 3, K = 5$, 
\item Number of repetitions, $n = 10$.
\end{itemize}

Moreover, the following variables:
\begin{itemize}
\item Dimensions of null space $d = 0, 5, 10, \dotsc, 50$,
\item Additional sire covariance scaling $c_A = 0.5, 1, 2$.
\end{itemize}

\medskip
\textit{Results:} \
For each combination of $\Sigma_A$ and $\Sigma_B$ structures
and the null space dimension and scaling variables $d$ and $p$
described above,
we compute the eigenvalues of the REML estimate $\sighat_{A,R}$ of
$\Sigma_A$. 

\begin{figure}[htbp]
  \centering
  \includegraphics[width=\textwidth]{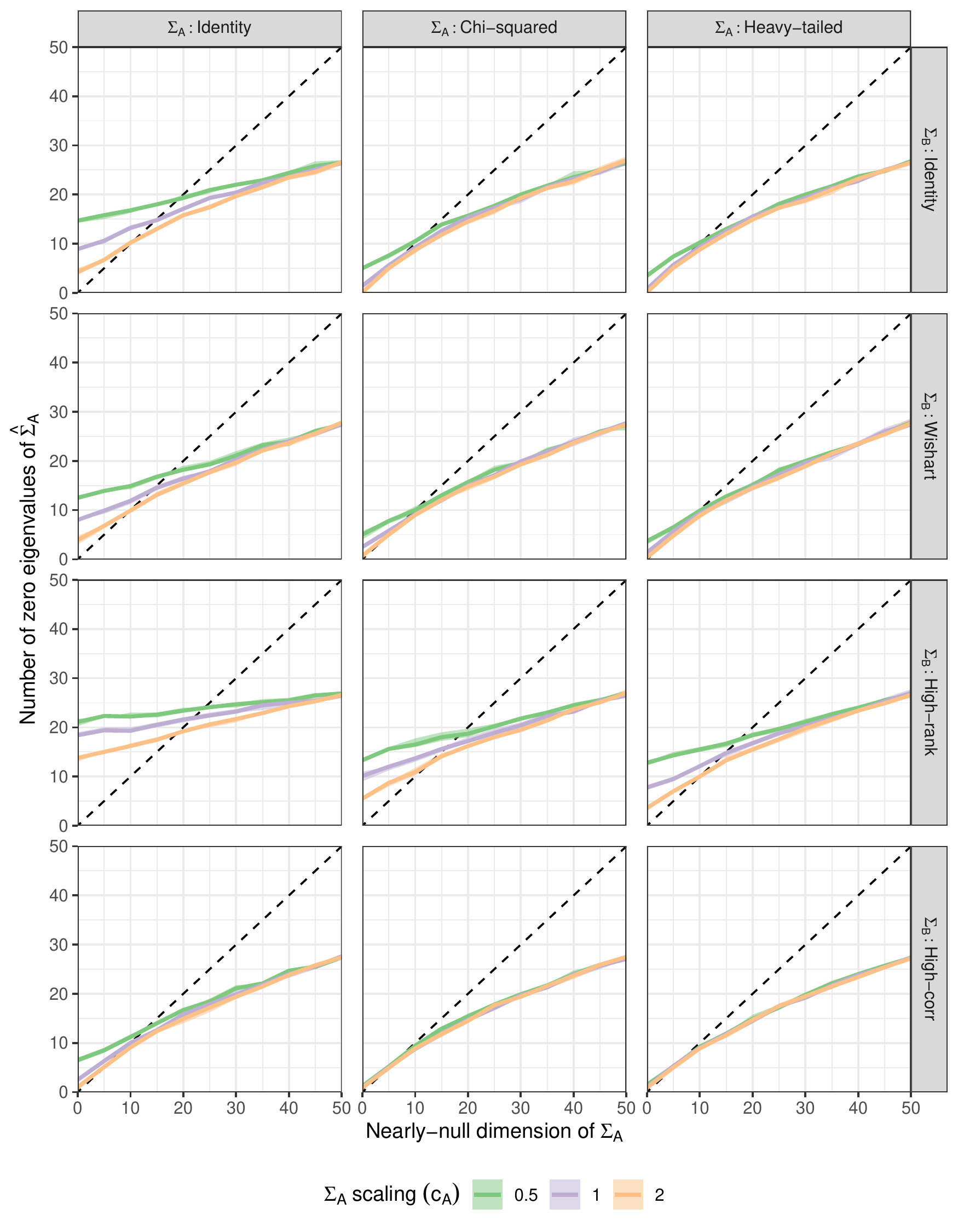}
  \caption{Number of zero eigenvalues of the REML estimate of $\Sigma_A$ plotted against the nearly-null dimenstion of $\Sigma_A$.
  The mean and inter-quartile range of 10 replicates are shown.}
  \label{fig:zero-evals}
\end{figure}

Let $\hat{d}$ be the number of eigenvalues of $\sighat_{A,R}$ that are
exactly zero.
Figure \ref{fig:zero-evals} plots $\hat{d}$ (vertical axis) against
$d$, the `true' null space dimension.
The solid line and ribbon show the mean and interquartile range
computed from the $n = 10$  replicates.
As a reference, the line $\hat{d} = d$ is shown on each plot with
black dashes. 

In all cases, the estimated nearly-null dimension tends to be
moderated: when $d$ is large, $\hat{d}$ is an underestimate, and
vice-versa when $d$ is small. 
Moreover, it seems that the structure of $B$ that is the main factor
governing the shape of the $\hat{d}$ curve.

\begin{figure}[htbp]
  \centering
  \includegraphics[width=\textwidth]{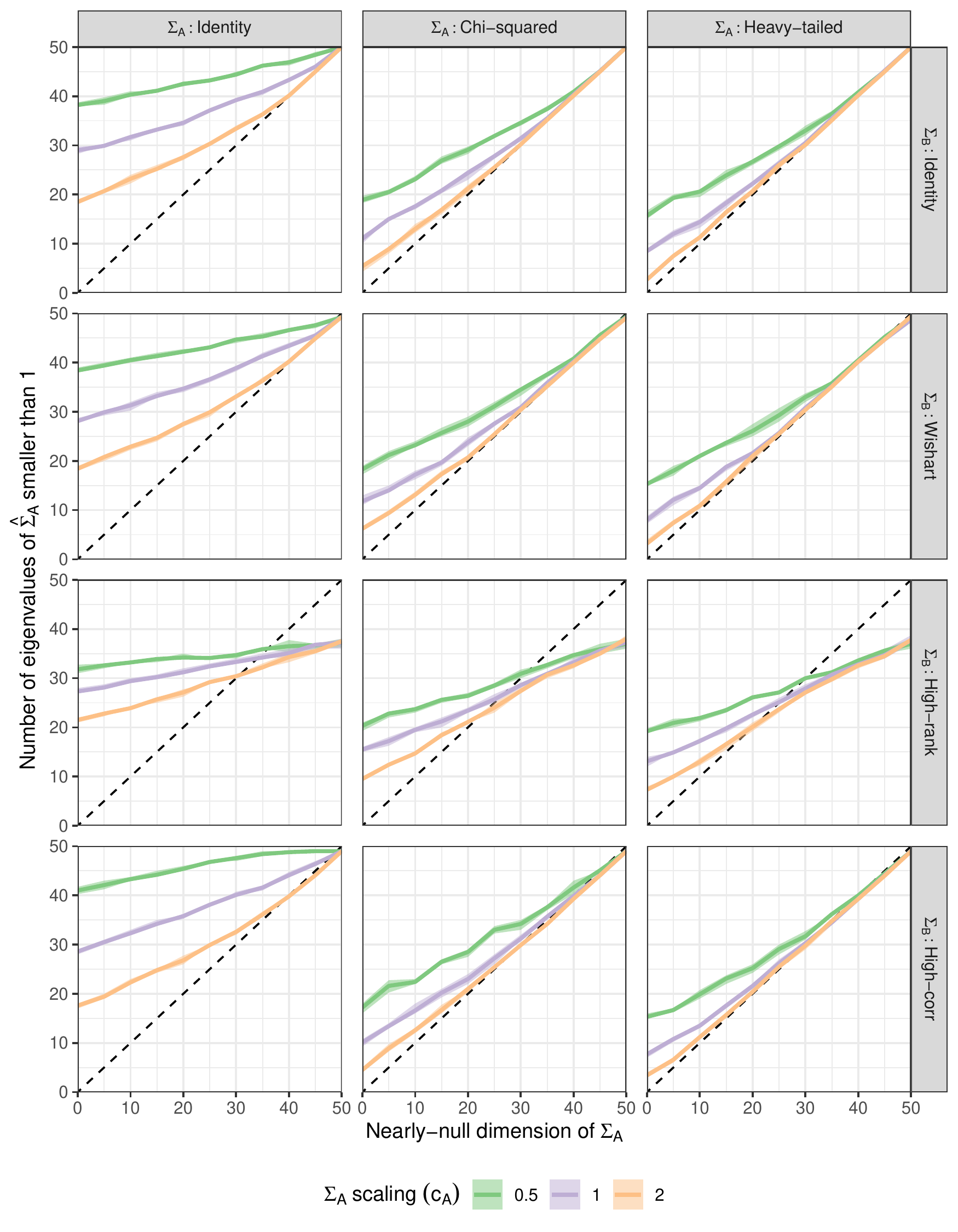}
  \caption{Number of eigenvalues of the REML estimate of $\Sigma_A$ smaller than $1$ plotted against the nearly-null dimension of $\Sigma_A$.
  The mean and inter-quartile range of 10 replicates are shown.}
  \label{fig:small-evals}
\end{figure}

We can perform similar analyses with different estimators of $d$.
Consider, for example
\begin{equation*}
  \hat{d}(\delta) = \# \{ i ~:~ \lambda_i \leq \delta \}.
\end{equation*}
\Cref{fig:small-evals} has the same features as \cref{fig:zero-evals}, but
instead displays the number of ``small'' eigenvalues of $A$,
defined
here as the number of eigenvalues with a value below $1$, i.e.
$\hat{d}(1)$.

This always exceeds the number of eigenvalues of $A$ that are exactly
zero, and so, as we might expect, it improves the under-estimates at
the expense of the over-estimates. 

This estimator does particularly well when $B$ has well-separated
eigenvalues - that is in the high-correlation and identity cases
and when the eigenvalues of $A$ are larger (chi-squared and heavy-tailed cases).

\subsection{Bias in \texorpdfstring{$\lambda_{1,R}$}{lambda(1,R)}
}
\label{sec:bias-}

Again we allow the dimension $d$ of the null space of $\Sigma_A$ to
vary, but now turn attention to biases in the REML estimate of the
largest eigenvalue.

The previous simulation setting is used, except that the following
structures are used for $\Sigma_A$:

\begin{equation}
  \Sigma_A =
   \begin{cases}
     c_A \diag( \bo_{p-d}, \bz_d)   & \text{Identity} \\
     c_A \diag( (X_i)_1^{p-d}, \bz_d) \qquad \quad X_i \stackrel{\rm
       iid}{\sim} \chi_5^2/5  & \text{Chi-squared(*)} \\
     c_A \diag( 5, \bo_{p-d-1}, \bz_d)   & \text{Spiked}.
   \end{cases}
\end{equation}
(*): the variates $X_i$ are held the same across replications to
reduce variation.

\begin{figure}[htbp]
  \centering
  \includegraphics[width=\textwidth]{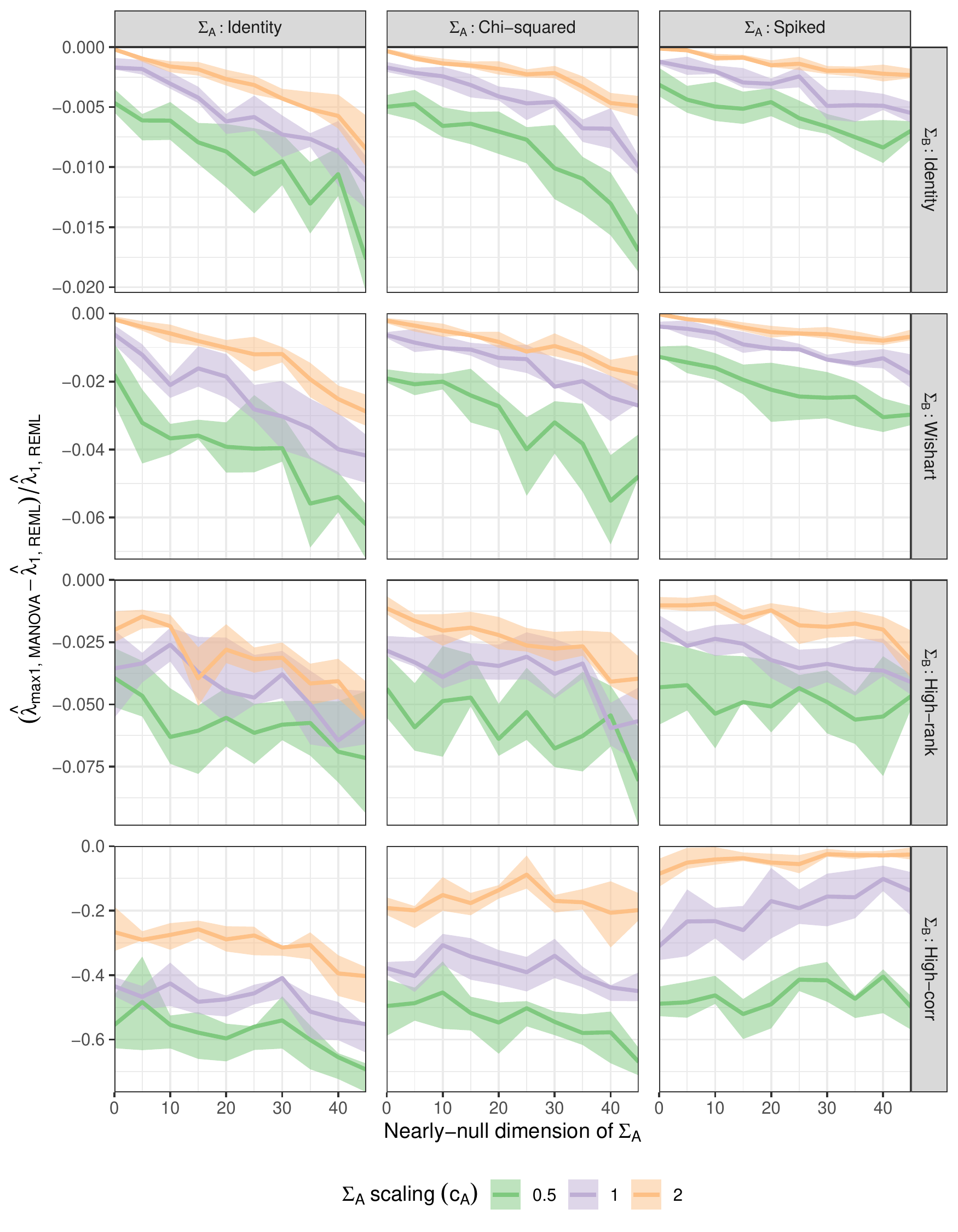}
  \caption{Relative difference between the largest eigenvalue of the MANOVA and REML estimate of $\Sigma_A$ for various choices of $\Sigma_A$ and $\Sigma_B$. $\Sigma_E = I$ throughout.
  The mean and inter-quartile range of 10 replicates are shown.}
  \label{fig:bias-largest}
\end{figure}

We focus on the top eigenvalue of the MANOVA and REML estimates of the
sire covariance $\Sigma_A$, namely $\lambda_{1,M}$ and $\lambda_{1,R}$
respectively across the settings, with $R=10$ replications.
Figure \ref{fig:bias-largest} shows the behavior of the relative
differences $(\lambda_{1,M} -\lambda_{1,R})/\lambda_{1,R}$.

\begin{itemize}
\item Empirically, it always happens that $\lambda_{1,R} \geq
\lambda_{1,M}$,
\item The size of the relative differences $(\lambda_{1,M} -
\lambda_{1,R})/\lambda_{1,R}$ depends strongly on $\Sigma_B$, being
much larger for $\Sigma_B =$ `highcorr' (which has a single large
spike eigenvalue) than for the other cases, where the difference is
smaller, but not insignificant,
\item the relative differences also tend to increase in magnitude with the null space
dimension $d$.
\item the relative differences also tend to vary inversely with the sire
  variance scaling $c_A$.  
\end{itemize}

\begin{figure}[htbp]
  \centering
  \includegraphics[width=\textwidth]{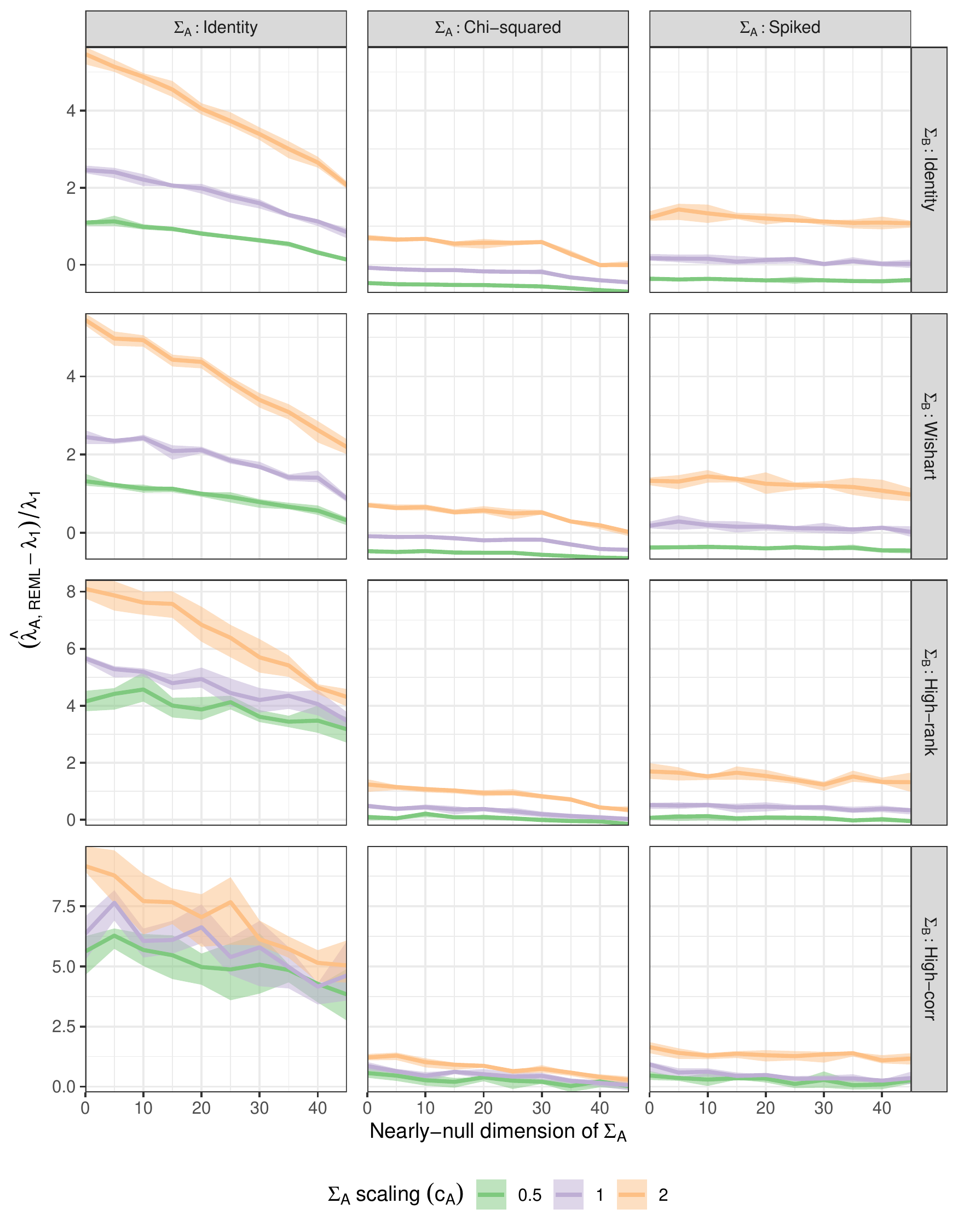}
  \caption{Relative bias in the largest eigenvalue of the REML estimate of $\Sigma_A$ for various choices of $\Sigma_A$ and $\Sigma_B$. $\Sigma_E = I$ throughout.
  The mean and inter-quartile range of 10 replicates are shown.}
  \label{fig:bias-to-truth}
\end{figure}

We can also investigate the bias in $\lambda_{1,R}$ by plotting the
relative difference $(\lambda_{1,R} - \lambda_1)/\lambda_1$
where $\lambda_1$ is the largest eigenvalue of the true underlying
covariance matrix $\Sigma_A$, Figure \ref{fig:bias-to-truth}.

\bigskip
We may make the following remarks about the upward bias of the REML
estimate relative to 
the truth: $(\lambda_{1,R} - \lambda_1)/\lambda_1$
\begin{itemize}
\item it depends on the structure, rank and magnitude of both $\Sigma_A$ and
  $\Sigma_B$ 
\item it is particularly large when the nonzero eigenvalues of $\Sigma_A$
  are all the same (`identity'),
  though in this case the eigenvalues of $\Sigma_A$ are smaller than for `chisquared' and `spiked',
\item it is smallest in the high correlation case
 $\Sigma_B$.
\end{itemize}





\section{Discussion}
\label{sec:conclusions}

We have shown that the Calvin-Dykstra convex duality algorithm renders
REML feasible for high-$p$ phenomics in the very specific setting of
balanced half-sib designs.\footnote{\citet{calv93} proposed an extension of the Calvin-Dykstra algorithm
to unbalanced half-sib designs, using the EM algorithm to impute
missing observations.
In current work we are implementing this algorithm in order to apply
it to some publicly available datasets that are beyond current
capabilities of more generic REML solvers.}

While the balanced observation assumption in this paper is unrealistic
in practice, it does allow initial demonstration of statistical
estimation phenomena for larqe $p$ that will likely carry over to
unbalanced settings and more complex designs.
We have seen examples of two such phenomena in the simulations
conducted here.

First there are significant biases in estimating the dimension $d$ of
a null space in the genetic covariance by simply counting the number
of zero REML eigenvalues: it is typically biased high when $d/p$ is small and
biased low when $d/p$ is large.
Second, there are significant biases of overestimation in the REML
estimates of the largest eigenvalues of genetic covariance matrices.

Both these high dimensional biases are not unexpected if one recalls
similar phenomena (eigenvalue spreading and biases) that have been
observed and substantiated by asymptotic approximation in two simpler
settings: estimation of a single $p$-dimensional covariance matrix
based on i.i.d. vector observations (reviewed in
\cite{jopa18}), and MANOVA estimates in
the present high dimensional variance component models
\citep{fan_eigenvalue_2019,fajo22,fan_spiked_2018}.
It is a natural topic for future research to look for greater
understanding of REML in high-$p$ settings, for example via asymptotic
approximations.

In summary, the simulations in this work, even with their
limitation to specific settings in special balanced designs, already
indicate that considerable caution will be needed in interpreting REML
estimates in phenome-wide studies of genetic variation.

\bibliographystyle{plainnat}
\bibliography{REML_Sim,JacBib3,mezey2005}

\begin{thebibliography}{51}
\providecommand{\natexlab}[1]{#1}
\providecommand{\url}[1]{\texttt{#1}}
\expandafter\ifx\csname urlstyle\endcsname\relax
  \providecommand{\doi}[1]{doi: #1}\else
  \providecommand{\doi}{doi: \begingroup \urlstyle{rm}\Url}\fi

\bibitem[Agrawal and Stinchcombe(2009)]{agrawal_how_2009}
Aneil~F Agrawal and John~R Stinchcombe.
\newblock How much do genetic covariances alter the rate of adaptation?
\newblock \emph{Proceedings of the Royal Society B: Biological Sciences},
  276\penalty0 (1659):\penalty0 1183--1191, March 2009.
\newblock ISSN 0962-8452, 1471-2954.
\newblock \doi{10.1098/rspb.2008.1671}.
\newblock URL
  \url{https://royalsocietypublishing.org/doi/10.1098/rspb.2008.1671}.

\bibitem[Amemiya(1985)]{amem85}
Yasuo Amemiya.
\newblock What should be done when an estimated between-group covariance matrix
  is not nonnegative definite?
\newblock \emph{The American Statistician}, 39\penalty0 (2):\penalty0 112--117,
  1985.

\bibitem[Blows and McGuigan(2015)]{blows_distribution_2015}
Mark~W. Blows and Katrina McGuigan.
\newblock The distribution of genetic variance across phenotypic space and the
  response to selection.
\newblock \emph{Molecular Ecology}, 24\penalty0 (9):\penalty0 2056--2072, May
  2015.
\newblock ISSN 09621083.
\newblock \doi{10.1111/mec.13023}.
\newblock URL \url{https://onlinelibrary.wiley.com/doi/10.1111/mec.13023}.

\bibitem[Blows et~al.(2015)Blows, Allen, Collet, Chenoweth, and
  McGuigan]{blows_phenome-wide_2015}
Mark~W. Blows, Scott~L. Allen, Julie~M. Collet, Stephen~F. Chenoweth, and
  Katrina McGuigan.
\newblock The {Phenome}-{Wide} {Distribution} of {Genetic} {Variance}.
\newblock \emph{The American Naturalist}, 186\penalty0 (1):\penalty0 15--30,
  July 2015.
\newblock ISSN 0003-0147, 1537-5323.
\newblock \doi{10.1086/681645}.
\newblock URL \url{https://www.journals.uchicago.edu/doi/10.1086/681645}.

\bibitem[Bégin and Roff(2004)]{begin_micro-_2004}
Mattieu Bégin and Derek~A. Roff.
\newblock From micro- to macroevolution through quantitative genetic variation:
  positive evidence from field crickets.
\newblock \emph{Evolution}, 58\penalty0 (10):\penalty0 2287--2304, October
  2004.
\newblock ISSN 0014-3820, 1558-5646.
\newblock \doi{10.1111/j.0014-3820.2004.tb01604.x}.
\newblock URL
  \url{https://onlinelibrary.wiley.com/doi/10.1111/j.0014-3820.2004.tb01604.x}.

\bibitem[Calvin(1993)]{calv93}
James~A. Calvin.
\newblock {REML} estimation in unbalanced multivariate variance components
  models using an em algorithm.
\newblock \emph{Biometrics}, 49\penalty0 (3):\penalty0 691--701, 1993.
\newblock ISSN 0006341X, 15410420.
\newblock URL \url{http://www.jstor.org/stable/2532190}.

\bibitem[Calvin and Dykstra(1991)]{cady91}
James~A. Calvin and Richard~L. Dykstra.
\newblock Maximum likelihood estimation of a set of covariance matrices under
  {L}\"{o}wner order restrictions with applications to balanced multivariate
  variance components models.
\newblock \emph{Ann. Statist.}, 19\penalty0 (2):\penalty0 850--869, 1991.

\bibitem[Caruso et~al.(2005)Caruso, Maherali, Mikulyuk, Carlson, and
  Jackson]{caruso_genetic_2005}
Christina~M. Caruso, Hafiz Maherali, Alison Mikulyuk, Kjarstin Carlson, and
  Robert~B. Jackson.
\newblock Genetic variance and covariance for physiological traits in
  lobelia:are there constraints on adaptive evolution?
\newblock \emph{Evolution}, 59\penalty0 (4):\penalty0 826--837, April 2005.
\newblock ISSN 0014-3820, 1558-5646.
\newblock \doi{10.1111/j.0014-3820.2005.tb01756.x}.
\newblock URL
  \url{https://onlinelibrary.wiley.com/doi/10.1111/j.0014-3820.2005.tb01756.x}.

\bibitem[Fan and Johnstone(2019)]{fan_eigenvalue_2019}
Zhou Fan and Iain~M. Johnstone.
\newblock Eigenvalue distributions of variance components estimators in
  high-dimensional random effects models.
\newblock \emph{The Annals of Statistics}, 47\penalty0 (5), October 2019.
\newblock ISSN 0090-5364.
\newblock \doi{10.1214/18-AOS1767}.
\newblock URL
  \url{https://projecteuclid.org/journals/annals-of-statistics/volume-47/issue-5/Eigenvalue-distributions-of-variance-components-estimators-in-high-dimensional-random/10.1214/18-AOS1767.full}.

\bibitem[Fan and Johnstone(2022)]{fajo22}
Zhou Fan and Iain~M. Johnstone.
\newblock {Tracy–Widom at each edge of real covariance and MANOVA
  estimators}.
\newblock \emph{The Annals of Applied Probability}, 32\penalty0 (4):\penalty0
  2967 -- 3003, 2022.
\newblock \doi{10.1214/21-AAP1754}.
\newblock URL \url{https://doi.org/10.1214/21-AAP1754}.

\bibitem[Fan et~al.(2018)Fan, Johnstone, and Sun]{fan_spiked_2018}
Zhou Fan, Iain~M. Johnstone, and Yi~Sun.
\newblock Spiked covariances and principal components analysis in
  high-dimensional random effects models, June 2018.
\newblock URL \url{http://arxiv.org/abs/1806.09529}.
\newblock arXiv:1806.09529 [math, stat].

\bibitem[Freimer and Sabatti(2003)]{freimer_human_2003}
Nelson Freimer and Chiara Sabatti.
\newblock The {Human} {Phenome} {Project}.
\newblock \emph{Nature Genetics}, 34\penalty0 (1):\penalty0 15--21, May 2003.
\newblock ISSN 1061-4036, 1546-1718.
\newblock \doi{10.1038/ng0503-15}.
\newblock URL \url{http://www.nature.com/articles/ng0503-15}.

\bibitem[Furbank and Tester(2011)]{furbank_phenomicstechnologies_2011}
Robert~T Furbank and Mark Tester.
\newblock Phenomics–technologies to relieve the phenotyping bottleneck.
\newblock \emph{Trends in plant science}, 16\penalty0 (12):\penalty0 635--644,
  2011.
\newblock Publisher: Elsevier.

\bibitem[Garcia et~al.(2014)Garcia, Hingst-Zaher, Cerqueira, and
  Marroig]{garcia_quantitative_2014}
Guilherme Garcia, Erika Hingst-Zaher, Rui Cerqueira, and Gabriel Marroig.
\newblock Quantitative {Genetics} and {Modularity} in {Cranial} and
  {Mandibular} {Morphology} of {Calomys} expulsus.
\newblock \emph{Evolutionary Biology}, 41\penalty0 (4):\penalty0 619--636,
  December 2014.
\newblock ISSN 0071-3260, 1934-2845.
\newblock \doi{10.1007/s11692-014-9293-4}.
\newblock URL \url{http://link.springer.com/10.1007/s11692-014-9293-4}.

\bibitem[Gaydos et~al.(2013)Gaydos, Heckman, Kirkpatrick, Stinchcombe, Schmitt,
  Kingsolver, and Marron]{gaydos_visualizing_2013}
Travis~L. Gaydos, Nancy~E. Heckman, Mark Kirkpatrick, J.~R. Stinchcombe,
  Johanna Schmitt, Joel Kingsolver, and J.~S. Marron.
\newblock Visualizing genetic constraints.
\newblock \emph{The Annals of Applied Statistics}, 7\penalty0 (2), June 2013.
\newblock ISSN 1932-6157.
\newblock \doi{10.1214/12-AOAS603}.
\newblock URL
  \url{https://projecteuclid.org/journals/annals-of-applied-statistics/volume-7/issue-2/Visualizing-genetic-constraints/10.1214/12-AOAS603.full}.

\bibitem[Gomulkiewicz and Houle(2009)]{gomulkiewicz_demographic_2009}
Richard Gomulkiewicz and David Houle.
\newblock Demographic and {Genetic} {Constraints} on {Evolution}.
\newblock \emph{The American Naturalist}, 174\penalty0 (6):\penalty0
  E218--E229, December 2009.
\newblock ISSN 0003-0147, 1537-5323.
\newblock \doi{10.1086/645086}.
\newblock URL \url{https://www.journals.uchicago.edu/doi/10.1086/645086}.

\bibitem[Gosden and Chenoweth(2014)]{gosden_evolutionary_2014}
Thomas~P. Gosden and Stephen~F. Chenoweth.
\newblock The evolutionary stability of cross-sex, cross-trait genetic
  covariances: evolutionary stability of cross-sex, cross-trait genetic
  covariances.
\newblock \emph{Evolution}, 68\penalty0 (6):\penalty0 1687--1697, June 2014.
\newblock ISSN 00143820.
\newblock \doi{10.1111/evo.12398}.
\newblock URL \url{https://onlinelibrary.wiley.com/doi/10.1111/evo.12398}.

\bibitem[Haleem et~al.(2021)Haleem, Javaid, Singh, Suman, and
  Rab]{haleem_biosensors_2021}
Abid Haleem, Mohd Javaid, Ravi~Pratap Singh, Rajiv Suman, and Shanay Rab.
\newblock Biosensors applications in medical field: {A} brief review.
\newblock \emph{Sensors International}, 2:\penalty0 100100, 2021.
\newblock ISSN 26663511.
\newblock \doi{10.1016/j.sintl.2021.100100}.
\newblock URL
  \url{https://linkinghub.elsevier.com/retrieve/pii/S2666351121000218}.

\bibitem[Hayes and Hill(1981)]{hayes_modification_1981}
J.~F. Hayes and W.~G. Hill.
\newblock Modification of {Estimates} of {Parameters} in the {Construction} of
  {Genetic} {Selection} {Indices} ('{Bending}').
\newblock \emph{Biometrics}, 37\penalty0 (3):\penalty0 483, September 1981.
\newblock ISSN 0006341X.
\newblock \doi{10.2307/2530561}.
\newblock URL \url{https://www.jstor.org/stable/2530561?origin=crossref}.

\bibitem[Hill and Thompson(1978)]{hill_probabilities_1978}
W.~G. Hill and R.~Thompson.
\newblock Probabilities of {Non}-{Positive} {Definite} between-{Group} or
  {Genetic} {Covariance} {Matrices}.
\newblock \emph{Biometrics}, 34\penalty0 (3):\penalty0 429, September 1978.
\newblock ISSN 0006341X.
\newblock \doi{10.2307/2530605}.
\newblock URL \url{https://www.jstor.org/stable/2530605?origin=crossref}.

\bibitem[Hill and Kirkpatrick(2010)]{hill_what_2010}
William~G. Hill and Mark Kirkpatrick.
\newblock What {Animal} {Breeding} {Has} {Taught} {Us} about {Evolution}.
\newblock \emph{Annual Review of Ecology, Evolution, and Systematics},
  41\penalty0 (1):\penalty0 1--19, December 2010.
\newblock ISSN 1543-592X, 1545-2069.
\newblock \doi{10.1146/annurev-ecolsys-102209-144728}.
\newblock URL
  \url{https://www.annualreviews.org/doi/10.1146/annurev-ecolsys-102209-144728}.

\bibitem[Hine et~al.(2014)Hine, McGuigan, and Blows]{hine_evolutionary_2014}
Emma Hine, Katrina McGuigan, and Mark~W. Blows.
\newblock Evolutionary {Constraints} in {High}-{Dimensional} {Trait} {Sets}.
\newblock \emph{The American Naturalist}, 184\penalty0 (1):\penalty0 119--131,
  July 2014.
\newblock ISSN 0003-0147, 1537-5323.
\newblock \doi{10.1086/676504}.
\newblock URL \url{https://www.journals.uchicago.edu/doi/10.1086/676504}.

\bibitem[Hine et~al.(2018)Hine, Runcie, McGuigan, and Blows]{hine_uneven_2018}
Emma Hine, Daniel~E Runcie, Katrina McGuigan, and Mark~W Blows.
\newblock Uneven {Distribution} of {Mutational} {Variance} {Across} the
  {Transcriptome} of \textit{{Drosophila} serrata} {Revealed} by
  {High}-{Dimensional} {Analysis} of {Gene} {Expression}.
\newblock \emph{Genetics}, 209\penalty0 (4):\penalty0 1319--1328, August 2018.
\newblock ISSN 1943-2631.
\newblock \doi{10.1534/genetics.118.300757}.
\newblock URL
  \url{https://academic.oup.com/genetics/article/209/4/1319/5931002}.

\bibitem[Houle and Meyer(2015)]{houle_estimating_2015}
D.~Houle and K.~Meyer.
\newblock Estimating sampling error of evolutionary statistics based on genetic
  covariance matrices using maximum likelihood.
\newblock \emph{Journal of Evolutionary Biology}, 28\penalty0 (8):\penalty0
  1542--1549, August 2015.
\newblock ISSN 1010-061X, 1420-9101.
\newblock \doi{10.1111/jeb.12674}.
\newblock URL \url{https://onlinelibrary.wiley.com/doi/10.1111/jeb.12674}.

\bibitem[Houle(2010)]{houle_numbering_2010}
David Houle.
\newblock Numbering the hairs on our heads: {The} shared challenge and promise
  of phenomics.
\newblock \emph{Proceedings of the National Academy of Sciences}, 107\penalty0
  (suppl\_1):\penalty0 1793--1799, January 2010.
\newblock ISSN 0027-8424, 1091-6490.
\newblock \doi{10.1073/pnas.0906195106}.
\newblock URL \url{https://pnas.org/doi/full/10.1073/pnas.0906195106}.

\bibitem[Houle et~al.(2010)Houle, Govindaraju, and
  Omholt]{houle_phenomics_2010}
David Houle, Diddahally~R. Govindaraju, and Stig Omholt.
\newblock Phenomics: the next challenge.
\newblock \emph{Nature Reviews Genetics}, 11\penalty0 (12):\penalty0 855--866,
  December 2010.
\newblock ISSN 1471-0056, 1471-0064.
\newblock \doi{10.1038/nrg2897}.
\newblock URL \url{http://www.nature.com/articles/nrg2897}.

\bibitem[Innocenti and Chenoweth(2013)]{innocenti_interspecific_2013}
Paolo Innocenti and Stephen~F. Chenoweth.
\newblock Interspecific {Divergence} of {Transcription} {Networks} along
  {Lines} of {Genetic} {Variance} in {Drosophila}: {Dimensionality},
  {Evolvability}, and {Constraint}.
\newblock \emph{Molecular Biology and Evolution}, 30\penalty0 (6):\penalty0
  1358--1367, June 2013.
\newblock ISSN 1537-1719, 0737-4038.
\newblock \doi{10.1093/molbev/mst047}.
\newblock URL
  \url{https://academic.oup.com/mbe/article-lookup/doi/10.1093/molbev/mst047}.

\bibitem[Jin et~al.(2020)Jin, Wilson, Beck, Nelson, Brownridge, Harrison,
  Djukovic, Raftery, Brem, Yu, Drton, Shojaie, Kapahi, and
  Promislow]{jin_genetic_2020}
Kelly Jin, Kenneth~A. Wilson, Jennifer~N. Beck, Christopher~S. Nelson,
  George~W. Brownridge, Benjamin~R. Harrison, Danijel Djukovic, Daniel Raftery,
  Rachel~B. Brem, Shiqing Yu, Mathias Drton, Ali Shojaie, Pankaj Kapahi, and
  Daniel Promislow.
\newblock Genetic and metabolomic architecture of variation in diet
  restriction-mediated lifespan extension in {Drosophila}.
\newblock \emph{PLOS Genetics}, 16\penalty0 (7):\penalty0 e1008835, July 2020.
\newblock ISSN 1553-7404.
\newblock \doi{10.1371/journal.pgen.1008835}.
\newblock URL \url{https://dx.plos.org/10.1371/journal.pgen.1008835}.

\bibitem[Johnstone and Paul(2018)]{jopa18}
I.~M. Johnstone and D.~Paul.
\newblock {PCA} in high dimensions: An orientation.
\newblock \emph{Proceedings of the IEEE}, 106\penalty0 (8):\penalty0
  1277--1292, Aug 2018.
\newblock ISSN 0018-9219.
\newblock \doi{10.1109/JPROC.2018.2846730}.

\bibitem[Johnstone(2001)]{john00c}
Iain~M. Johnstone.
\newblock On the distribution of the largest eigenvalue in principal components
  analysis.
\newblock \emph{Annals of Statistics}, 29:\penalty0 295--327, 2001.

\bibitem[Kirkpatrick(2009)]{kirkpatrick_patterns_2009}
Mark Kirkpatrick.
\newblock Patterns of quantitative genetic variation in multiple dimensions.
\newblock \emph{Genetica}, 136\penalty0 (2):\penalty0 271--284, June 2009.
\newblock ISSN 0016-6707, 1573-6857.
\newblock \doi{10.1007/s10709-008-9302-6}.
\newblock URL \url{http://link.springer.com/10.1007/s10709-008-9302-6}.

\bibitem[Lande(1980)]{lande_genetic_1980}
Russell Lande.
\newblock The genetic covariance between characters maintained by pleiotropic
  mutations.
\newblock \emph{Genetics}, 94\penalty0 (1):\penalty0 203--215, January 1980.
\newblock ISSN 1943-2631.
\newblock \doi{10.1093/genetics/94.1.203}.
\newblock URL \url{https://academic.oup.com/genetics/article/94/1/203/5993665}.

\bibitem[Lynch et~al.(1998)Lynch, Walsh, and {others}]{lynch_genetics_1998}
Michael Lynch, Bruce Walsh, and {others}.
\newblock \emph{Genetics and analysis of quantitative traits}, volume~1.
\newblock Sinauer Sunderland, MA, 1998.

\bibitem[Lürig et~al.(2021)Lürig, Donoughe, Svensson, Porto, and
  Tsuboi]{lurig_computer_2021}
Moritz~D. Lürig, Seth Donoughe, Erik~I. Svensson, Arthur Porto, and Masahito
  Tsuboi.
\newblock Computer {Vision}, {Machine} {Learning}, and the {Promise} of
  {Phenomics} in {Ecology} and {Evolutionary} {Biology}.
\newblock \emph{Frontiers in Ecology and Evolution}, 9:\penalty0 642774, April
  2021.
\newblock ISSN 2296-701X.
\newblock \doi{10.3389/fevo.2021.642774}.
\newblock URL
  \url{https://www.frontiersin.org/articles/10.3389/fevo.2021.642774/full}.

\bibitem[Mar\v{c}enko and Pastur(1967)]{mapa67}
V.~A. Mar\v{c}enko and L.~A. Pastur.
\newblock Distributions of eigenvalues of some sets of random matrices.
\newblock \emph{Math. USSR-Sb.}, 1:\penalty0 507--536, 1967.

\bibitem[McGlothlin et~al.(2022)McGlothlin, Kobiela, Wright, Kolbe, Losos, and
  Brodie]{mcglothlin_conservation_2022}
Joel~W. McGlothlin, Megan~E. Kobiela, Helen~V. Wright, Jason~J. Kolbe,
  Jonathan~B. Losos, and Edmund~D. Brodie.
\newblock Conservation and {Convergence} of {Genetic} {Architecture} in the
  {Adaptive} {Radiation} of \textit{{Anolis}} {Lizards}.
\newblock \emph{The American Naturalist}, pages E000--E000, September 2022.
\newblock ISSN 0003-0147, 1537-5323.
\newblock \doi{10.1086/721091}.
\newblock URL \url{https://www.journals.uchicago.edu/doi/10.1086/721091}.

\bibitem[Meyer(2019)]{meyer_bending_2019}
Karin Meyer.
\newblock “{Bending}” and beyond: {Better} estimates of quantitative
  genetic parameters?
\newblock \emph{Journal of Animal Breeding and Genetics}, 136\penalty0
  (4):\penalty0 243--251, July 2019.
\newblock ISSN 0931-2668, 1439-0388.
\newblock \doi{10.1111/jbg.12386}.
\newblock URL \url{https://onlinelibrary.wiley.com/doi/10.1111/jbg.12386}.

\bibitem[Mezey and Houle(2005)]{mezey_dimensionality_2005}
Jason~G Mezey and David Houle.
\newblock The dimensionality of genetic variation for wing shape in
  {Drosophila} melanogaster.
\newblock \emph{Evolution}, 59\penalty0 (5):\penalty0 1027--1038, 2005.
\newblock Publisher: Wiley Online Library.

\bibitem[Neethirajan(2017)]{neethirajan_recent_2017}
Suresh Neethirajan.
\newblock Recent advances in wearable sensors for animal health management.
\newblock \emph{Sensing and Bio-Sensing Research}, 12:\penalty0 15--29,
  February 2017.
\newblock ISSN 22141804.
\newblock \doi{10.1016/j.sbsr.2016.11.004}.
\newblock URL
  \url{https://linkinghub.elsevier.com/retrieve/pii/S2214180416301350}.

\bibitem[Runcie and Mukherjee(2013)]{runcie_dissecting_2013}
Daniel~E Runcie and Sayan Mukherjee.
\newblock Dissecting {High}-{Dimensional} {Phenotypes} with {Bayesian} {Sparse}
  {Factor} {Analysis} of {Genetic} {Covariance} {Matrices}.
\newblock \emph{Genetics}, 194\penalty0 (3):\penalty0 753--767, July 2013.
\newblock ISSN 1943-2631.
\newblock \doi{10.1534/genetics.113.151217}.
\newblock URL
  \url{https://academic.oup.com/genetics/article/194/3/753/6065385}.

\bibitem[Runcie et~al.(2021)Runcie, Qu, Cheng, and
  Crawford]{runcie_megalmm_2021}
Daniel~E. Runcie, Jiayi Qu, Hao Cheng, and Lorin Crawford.
\newblock {MegaLMM}: {Mega}-scale linear mixed models for genomic predictions
  with thousands of traits.
\newblock \emph{Genome Biology}, 22\penalty0 (1):\penalty0 213, December 2021.
\newblock ISSN 1474-760X.
\newblock \doi{10.1186/s13059-021-02416-w}.
\newblock URL
  \url{https://genomebiology.biomedcentral.com/articles/10.1186/s13059-021-02416-w}.

\bibitem[Saccenti et~al.(2011)Saccenti, Smilde, Westerhuis, and
  Hendriks]{saccenti_tracy-widom_2011}
Edoardo Saccenti, Age~K. Smilde, Johan~A. Westerhuis, and Margriet M. W.~B.
  Hendriks.
\newblock Tracy-{Widom} statistic for the largest eigenvalue of autoscaled real
  matrices: {TW} statistic for autoscaled real matrices.
\newblock \emph{Journal of Chemometrics}, 25\penalty0 (12):\penalty0 644--652,
  December 2011.
\newblock ISSN 08869383.
\newblock \doi{10.1002/cem.1411}.
\newblock URL \url{https://onlinelibrary.wiley.com/doi/10.1002/cem.1411}.

\bibitem[Schluter(1996)]{schluter_adaptive_1996}
Dolph Schluter.
\newblock Adaptive {Radiation} {Along} {Genetic} {Lines} of {Least}
  {Resistance}.
\newblock \emph{Evolution}, 50\penalty0 (5):\penalty0 1766, October 1996.
\newblock ISSN 00143820.
\newblock \doi{10.2307/2410734}.
\newblock URL \url{https://www.jstor.org/stable/2410734?origin=crossref}.

\bibitem[Schrag et~al.(2018)Schrag, Westhues, Schipprack, Seifert, Thiemann,
  Scholten, and Melchinger]{schrag_beyond_2018}
Tobias~A Schrag, Matthias Westhues, Wolfgang Schipprack, Felix Seifert,
  Alexander Thiemann, Stefan Scholten, and Albrecht~E Melchinger.
\newblock Beyond {Genomic} {Prediction}: {Combining} {Different} {Types} of
  \textit{omics} {Data} {Can} {Improve} {Prediction} of {Hybrid} {Performance}
  in {Maize}.
\newblock \emph{Genetics}, 208\penalty0 (4):\penalty0 1373--1385, April 2018.
\newblock ISSN 1943-2631.
\newblock \doi{10.1534/genetics.117.300374}.
\newblock URL
  \url{https://academic.oup.com/genetics/article/208/4/1373/6084222}.

\bibitem[Sharma et~al.(2021)Sharma, Badea, Tiwari, and
  Marty]{sharma_wearable_2021}
Atul Sharma, Mihaela Badea, Swapnil Tiwari, and Jean~Louis Marty.
\newblock Wearable {Biosensors}: {An} {Alternative} and {Practical} {Approach}
  in {Healthcare} and {Disease} {Monitoring}.
\newblock \emph{Molecules}, 26\penalty0 (3):\penalty0 748, February 2021.
\newblock ISSN 1420-3049.
\newblock \doi{10.3390/molecules26030748}.
\newblock URL \url{https://www.mdpi.com/1420-3049/26/3/748}.

\bibitem[Sztepanacz and Blows(2015)]{sztepanacz_dominance_2015}
Jacqueline~L Sztepanacz and Mark~W Blows.
\newblock Dominance {Genetic} {Variance} for {Traits} {Under} {Directional}
  {Selection} in \textit{{Drosophila} serrata}.
\newblock \emph{Genetics}, 200\penalty0 (1):\penalty0 371--384, May 2015.
\newblock ISSN 1943-2631.
\newblock \doi{10.1534/genetics.115.175489}.
\newblock URL
  \url{https://academic.oup.com/genetics/article/200/1/371/5936189}.

\bibitem[Sztepanacz and Blows(2017)]{sztepanacz_accounting_2017}
Jacqueline~L Sztepanacz and Mark~W Blows.
\newblock Accounting for {Sampling} {Error} in {Genetic} {Eigenvalues} {Using}
  {Random} {Matrix} {Theory}.
\newblock \emph{Genetics}, 206\penalty0 (3):\penalty0 1271--1284, July 2017.
\newblock ISSN 1943-2631.
\newblock \doi{10.1534/genetics.116.198606}.
\newblock URL
  \url{https://academic.oup.com/genetics/article/206/3/1271/6064212}.

\bibitem[Sztepanacz and Houle(2019)]{sztepanacz_crosssex_2019}
Jacqueline~L. Sztepanacz and David Houle.
\newblock Cross‐sex genetic covariances limit the evolvability of
  wing‐shape within and among species of \textit{{Drosophila}}.
\newblock \emph{Evolution}, 73\penalty0 (8):\penalty0 1617--1633, August 2019.
\newblock ISSN 0014-3820, 1558-5646.
\newblock \doi{10.1111/evo.13788}.
\newblock URL \url{https://onlinelibrary.wiley.com/doi/10.1111/evo.13788}.

\bibitem[Sztepanacz and Rundle(2012)]{sztepanacz_reduced_2012}
Jacqueline~L. Sztepanacz and Howard~D. Rundle.
\newblock Reduced genetic variance among high fitness individuals: inferring
  stabilizing selection on male sexual displays in \textit{drosophila serrata}:
  inferring stabilizing selection on sexual displays.
\newblock \emph{Evolution}, 66\penalty0 (10):\penalty0 3101--3110, October
  2012.
\newblock ISSN 00143820.
\newblock \doi{10.1111/j.1558-5646.2012.01658.x}.
\newblock URL
  \url{https://onlinelibrary.wiley.com/doi/10.1111/j.1558-5646.2012.01658.x}.

\bibitem[Walsh and Lynch(2018)]{walsh_evolution_2018}
Bruce Walsh and Michael Lynch.
\newblock \emph{Evolution and selection of quantitative traits}.
\newblock Oxford University Press, 2018.

\bibitem[Yao et~al.(2015)Yao, Zheng, and Bai]{yzb15}
Jianfeng Yao, Shurong Zheng, and Zhidong Bai.
\newblock \emph{Large sample covariance matrices and high-dimensional data
  analysis}, volume~39 of \emph{Cambridge Series in Statistical and
  Probabilistic Mathematics}.
\newblock Cambridge University Press, New York, 2015.
\newblock Chapter 10.

\end{thebibliography}

\end{document}